\documentclass{PoS}

\usepackage{graphicx}
\usepackage{amsmath}
\usepackage{amssymb}

\newcommand{\be}{\begin{equation}}
\newcommand{\ee}{\end{equation}}
\newcommand{\bea}{\begin{eqnarray}}
\newcommand{\eea}{\end{eqnarray}}
\newcommand{\bi}{\begin{itemize}}
\newcommand{\ei}{\end{itemize}}
\newcommand{\ben}{\begin{enumerate}}
\newcommand{\een}{\end{enumerate}}
\newcommand{\bt}{\begin{tabbing}}
\newcommand{\et}{\end{tabbing}}

\newcommand{\calO}{{\mathcal O}}

\newcommand{\crad}{\langle r^2 \rangle}
\newcommand{\pp}{{p^\prime}}
\newcommand{\bfp}{{\bf p}}
\newcommand{\bfpp}{{{\bf p}^\prime}}

\newcommand{\bfx}{{\bf x}}
\newcommand{\bfxp}{{{\bf x}^\prime}}
\newcommand{\bfxpp}{{{\bf x}^{\prime\prime}}}
\newcommand{\bfr}{{\bf r}}
\newcommand{\bfz}{{\bf 0}}
\newcommand{\dt}{{\Delta t}}
\newcommand{\dtp}{{\Delta t^\prime}}

\title{
   \begin{picture}(0,0)(0,0)%
   \put(350,75){\makebox(0,0)[l]{\textnormal{\normalsize KEK-CP-258}}}%
   \put(350,60){\makebox(0,0)[l]{\textnormal{\normalsize OU-HET-731-2011}}}%
   \end{picture}%
   Kaon semileptonic form factors in QCD with exact chiral symmetry
}

\ShortTitle{
Kaon semileptonic form factors in QCD with exact chiral symmetry
}

\author{
   JLQCD Collaboration: 
   \speaker{T.~Kaneko}$^{a,b}$\thanks{E-mail: takashi.kaneko@kek.jp}, 
   S.~Aoki$^c$, 
   G.~Cossu$^a$, 
   X.~Feng$^a$, 
   H.~Fukaya$^d$, 
   S.~Hashimoto$^{a,b}$, 
   J.~Noaki$^{a}$
   and
   T.~Onogi$^d$
   \\
   \\
   \\
   \llap{$^a$}
   High Energy Accelerator Research Organization (KEK),
   Ibaraki 305-0801, Japan 
   \\
   \llap{$^b$}
   School of High Energy Accelerator Science,
   The Graduate University for Advanced Studies (Sokendai),
   Ibaraki 305-0801, Japan
   \\ 
   \llap{$^c$}
   Graduate School of Pure and Applied Sciences, 
   University of Tsukuba, Ibaraki 305-8571, Japan
   \\
   \llap{$^d$}
   Department of Physics, Osaka University, 
   Toyonaka, Osaka 560-0043 Japan
}
   
\abstract{
We report on our calculation of the kaon semileptonic form factors
in $N_f\!=\!2\!+\!1$ lattice QCD. 
Chiral symmetry is exactly preserved by using the overlap quark action
for a straightforward comparison with chiral perturbation theory (ChPT).
We simulate three pion masses down to 290~MeV
at a single lattice spacing of 0.11 fm 
and at a strange quark mass very close to its physical value.
The form factors near zero momentum transfer 
are precisely calculated
by using the all-to-all propagator and twisted boundary conditions. 
We compare the normalizations and slopes of the form factors 
with ChPT and experiments.
}

\FullConference{ 
   The XXIX International Symposium on Lattice Field Theory - Lattice 2011\\
   July 10-16, 2011\\
   Squaw Valley, Lake Tahoe, California
}

\begin{document}


\section{Numerical simulations}

The matrix element of the $K\!\to\!\pi l \nu$ decays is parametrized 
by two form factors
\bea
   \langle \pi(p^\prime) | V_\mu | K(p) \rangle 
   & = & 
   (p+p^\prime)_\mu f_+(q^2) + (p-p^\prime)_\mu f_-(q^2)
   \hspace{3mm}
   (q^2=(p-p^\prime)^2).
   \label{eqn:intro:ME}
\eea
The normalization of the vector form factor $f_+(0)$ is an important quantity 
for a precise determination of a CKM matrix element $|V_{us}|$ and
the search for new physics.
For a reliable lattice calculation of $f_+(0)$,
we also examine the consistency of 
other information in the matrix element,
namely $f_-(0)$ and the form factors' shape, 
with chiral perturbation theory (ChPT) and experiments.

In this article, we report on our calculation of $f_{\{+,-\}}(q^2)$
in $N_f\!=\!2\!+\!1$ QCD. 
Chiral symmetry is exactly preserved by using the overlap quark action 
for a straightforward comparison with ChPT.
At a single lattice spacing $a\!=\!0.112(1)$~fm, 
we simulate three values of degenerate up and down quark masses 
$m_{ud}\!=\!0.015$, 0.035 and 0.050
that cover a range of the pion mass 290\,--\,540~MeV.
The strange quark mass is fixed to a single value $m_s\!=\!0.080$,
which is very close to its physical value $m_{s,\rm phys}\!=\!0.081$.
We choose a lattice size, $(L/a)^3 \!\times\! (T/a) = 16^3 \!\times\! 48$ 
or $24^3 \!\times\! 48$, depending on $m_{ud}$ 
in order to satisfy a condition $M_\pi L \! \gtrsim \! 4$ 
to control finite volume effects.
The statistics are 2,500 HMC trajectories 
at each combination of $m_{ud}$ and $m_s$.

We calculate two- and three-point functions
\bea
   C^P(\dt,\bfp)
   & = &
   \frac{a^4}{L^3 T}\sum_{\bfx,t} 
   \sum_{\bfxp}
   \langle 
      \calO_P(\bfxp,t+\dt) \calO_P^\dagger(\bfx,t) 
   \rangle,
   \label{eqn:sim_param:msn_corr_2pt}
   \\
   C_\mu^{PQ}(\dt,\dtp;\bfp,\bfpp)
   & = &
   \frac{a^4}{L^3 T}\sum_{\bfx,t} 
   \sum_{\bfxpp, \bfxp}
   \langle 
      \calO_Q(\bfxpp,t+\dt+\dtp) V_\mu(\bfxp,t+\dt) \calO_P^\dagger(\bfx,t) 
   \rangle
   \label{eqn:sim_param:msn_corr_3pt}
\eea
using an exponential smearing function $\phi(\bfr)\!=\!e^{-0.4|\bfr|}$ 
for the interpolating operator 
$\calO_P^\dagger(\bfx,t)
 = \sum_{\bfr} \phi(\bfr) \bar{q}(\bfx+\bfr,t) \gamma_5 q^\prime(\bfx,t)$
($P=\pi \mbox{\ or\ } K$).
We refer readers to Refs.~\cite{prev0,prev1}
for details on how to construct these correlators 
using the all-to-all propagator~\cite{A2A}.

In order to explore the most important kinematical region $q^2\!\sim\!0$, 
the meson momentum $\bfp^{(\prime)}$ is induced 
by employing the twisted boundary conditions (TBCs) \cite{TBC}
\bea
   q(\bfx+L\,\hat{k},t) = e^{i\theta}q(\bfx,t),
   \hspace{3mm}
   \bar{q}(\bfx+L\,\hat{k},t) = e^{-i\theta}\bar{q}(\bfx,t)
   \hspace{3mm} 
   (k=1,2,3),
   \label{eqn:intro:TBC}
\eea
where $\hat{k}$ is a unit vector in the $k$ direction.
We use a common twist angle $\theta$ in all the spatial directions
for simplicity.
Here we consider the $K^+ \! \to \! \pi^0 l \nu$ channel, 
and impose the TBCs for the up and strange quarks.
The periodic boundary condition is used for the spectator down quark. 
Our simulation parameters are summarized in Table~\ref{tbl:intro:param}.

\begin{table}[b]
\begin{center}
\caption{
   Simulation parameters. 
   We denote the bare quark masses in lattice units by $m_{\{ud,s\}}$.
}
\vspace{3mm}
\label{tbl:intro:param}
\begin{tabular}{lllll}
   \hline
   $m_{ud}$  & $m_s$  & lattice         & $M_\pi L$ & $\theta$
   \\ \hline
   0.050    & 0.080    & $16^3 \times 48$ & 4.9     & 0.00, 0.40, 0.96, 1.60
   \\
   0.035    & 0.080    & $16^3 \times 48$ & 4.1     & 0.00, 0.60, 1.28, 1.76
   \\
   0.015    & 0.080    & $24^3 \times 48$ & 4.2     & 0.00, 1.68, 2.64
   \\ \hline
\end{tabular}
\end{center}
\vspace{-5mm}
\end{table}


\begin{figure}[t]
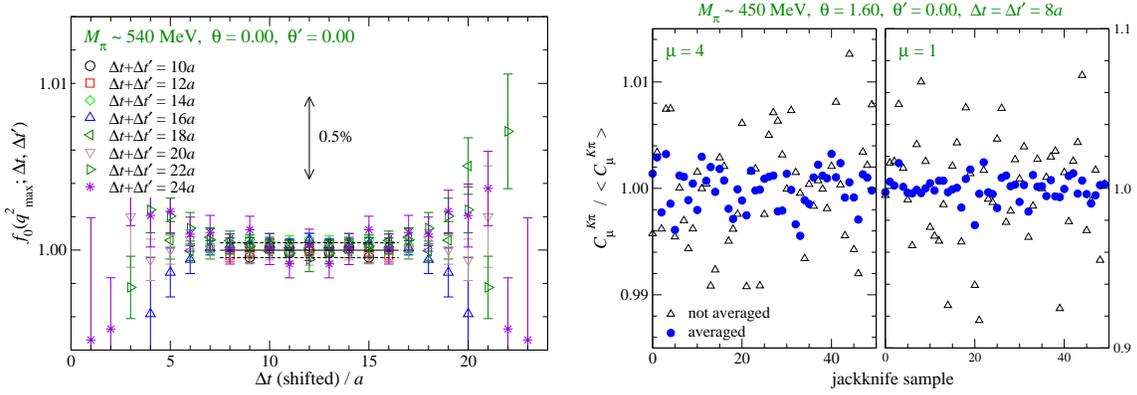

\begin{center}
   \includegraphics[angle=0,width=0.48\linewidth,clip]%
                   {kff_dble14_mud3_ms6_mval336_smr11.eps}
   \hspace{3mm}
   \includegraphics[angle=0,width=0.48\linewidth,clip]%
                   {msn_3pt_p-p-v4_mud2_ms6_mval226_tbc002_smr11.eps}
   \caption{
     Left panel: 
     effective value 
     $f_0(q^2_{\rm max};\dt,\dtp)\!=\!2 \sqrt{M_K M_\pi R(\dt,\dtp)}/(M_K+M_\pi)$
     obtained with different values of $\dt+\dtp$.
     Data are shifted in the horizontal direction
     so that the meson source and sink operators 
     are located at $T/4-(\dt+\dtp)/2$ and $T/4+(\dt+\dtp)/2$,
     respectively.
     %
     %
     Right two panels: 
     three-point function $C_4^{K\pi}$ for each jackknife sample. 
     We plot data normalized by the Monte Carlo average for $\mu\!=\!4$ and 1
     in each panel.
     Open and filled symbols are data with and without averaging over the 
     locations of the meson source, respectively.
   }
   \label{fig:drat:drat}
\end{center}
\vspace{-5mm}
\end{figure}

\section{Form factors at simulation points}
\label{sec:ff}

We calculate the scalar form factor 
$f_0(q^2)\!=\!f_+(q^2)+f_-(q^2)\,q^2/(M_K^2-M_\pi^2)$
at $q^2_{\rm max}\!=\!(M_K-M_\pi)^2$ 
from the following double ratio~\cite{kl3:drat:1}
\bea
   R(\dt, \dt^\prime) 
   & = & 
   \frac{C^{K \pi}_4(\dt,\dtp; \bfz, \bfz)
         C^{\pi K}_4(\dt,\dtp; \bfz, \bfz)}
        {C^{K K}_4(\dt,\dtp; \bfz, \bfz)
         C^{\pi \pi}_4(\dt,\dtp; \bfz, \bfz)}
   \xrightarrow[\dt, \dtp \to \infty]{} 
   \frac{(M_K+M_\pi)^2}{4 M_K M_\pi} f_0(q^2_{\rm max})^2.
   \label{eqn:drat:drat1}
\eea
The form factors $f_{\{+,0\}}(q^2)$ at $q < q^2_{\rm max}$ 
are calculated from~\cite{kl3:drat:1,kl3:drat:2}
\bea
   \tilde{R} 
   & = & 
   \frac{C^{K \pi}_4(\dt,\dtp; \bfp, \bfpp)
         C^{K}(\dt, \bfz)\, C^{\pi}(\dtp, \bfz)}
        {C^{K \pi}_4(\dt, \dtp; \bfz, \bfz)
         C^{K}(\dt,\bfp)\, C^{\pi}(\dtp, \bfpp)}
   %
   \to 
   \left\{
      \frac{E_K+E_\pi^\prime}{M_K+M_\pi} 
    + \frac{E_K-E_\pi^\prime}{M_K+M_\pi} \xi(q^2)
   \right\}
   \frac{f_+(q^2)}{f_0(q^2_{\rm max})},
   \hspace{9mm}
   \label{eqn:drat:drat2}
   \\[1mm]
   R_k 
   & = & 
   \frac{C^{K \pi}_k(\dt,\dtp; \bfp, \bfpp)
         C^{KK}_4(\dt, \dtp; \bfp, \bfpp)}
        {C^{K \pi}_4(\dt, \dtp; \bfp, \bfpp)
         C^{KK}_k(\dt, \dtp; \bfp, \bfpp)}
   \to 
   \frac{2p_k}{(p+\pp)_k}
   \frac{E_K+E_K^\prime}{(E_K-E_\pi^\prime)\xi(q^2)-E_K-E_\pi^\prime},
   \label{eqn:drat:drat3}
\eea
where $E_P^{(\prime)}$ ($P=\pi$ or $K$) 
represents the energy of the meson $P$ with the momentum $\bfp^{(\prime)}$.
Note that we can convert $f_+(q^2)$ to $f_0(q^2)$ (and vice versa)
using the the ratio $\xi(q^2)=f_-(q^2)/f_+(q^2)$,
except at $q^2_{\rm max}$ where $\tilde{R}$ and $R_k$ have 
no sensitivity to $\xi(q^2)$.

The use of the all-to-all propagator greatly helps us identify 
the plateaux of these ratios.
In contrast to previous studies with $\dt+\dtp$ kept fixed, 
we can take arbitrary combinations of $\dt$ and $\dtp$.
As shown in Fig.~\ref{fig:drat:drat}, 
the effective values of $f_0(q^2_{\rm max})$ 
with different values of $\dt+\dtp$ exhibit good consistency, 
which gives us confidence about our determination with sub-percent accuracy.
%

Another important advantage with the all-to-all propagator is that 
we can remarkably improve the statistical accuracy of 
meson correlators (and hence their ratios)
by averaging over the locations of the meson source $(\bfx,t)$
in Eqs.~(\ref{eqn:sim_param:msn_corr_2pt}) and 
(\ref{eqn:sim_param:msn_corr_3pt}). 
As shown in the right two panels of Fig.~\ref{fig:drat:drat}, 
the statistical fluctuation of $C_\mu^{K\pi}$ ($\mu\!=\!4,1$)
is reduced by about a factor of 3.
This averaging enables us to 
achieve $\lesssim1$\,\% accuracy for $f_{\{+,0\}}(q^2)$,
and 10\,--\,30\,\% for $\xi(q^2)$.


\section{$q^2$ dependence}

\begin{figure}[t]
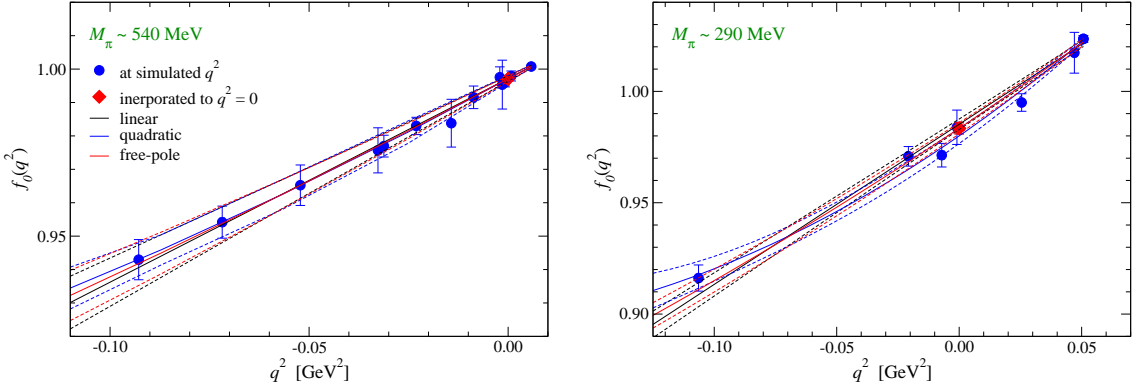

\begin{center}
   \includegraphics[angle=0,width=0.48\linewidth,clip]%
                   {f0_vs_q2_mud0050_ms0080.eps}
   \hspace{3mm}
   \includegraphics[angle=0,width=0.48\linewidth,clip]%
                   {f0_vs_q2_mud0015_ms0080.eps}
   \caption{
      Scalar form factor $f_0(q^2)$ as a function of $q^2$. 
      Left and right panels show data at our heaviest and lightest pion masses,
      respectively.
      We also plot interpolations to $q^2\!=\!0$ 
      with various parametrization forms 
      together with $f_0(0)$ obtained using the pole ansatz (diamond).
   }
   \label{fig:q2_dep:f0}
\end{center}
\vspace{-5mm}
\end{figure}

\begin{figure}[b]
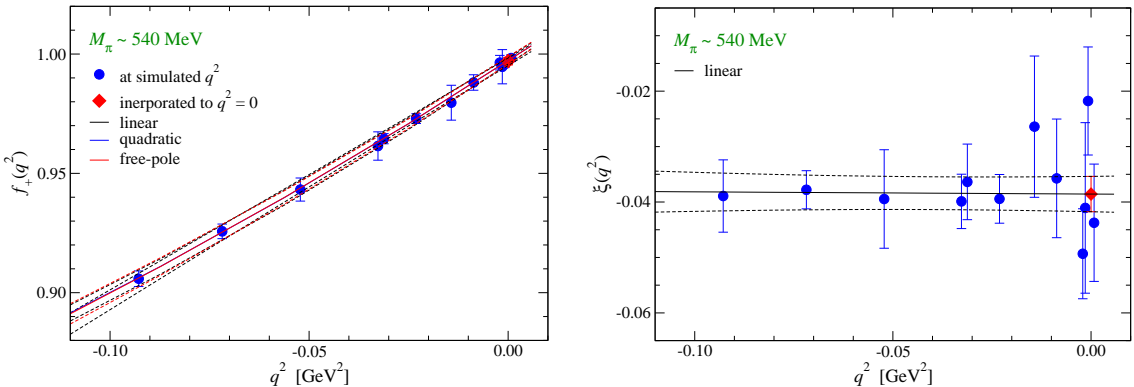

\begin{center}
   \includegraphics[angle=0,width=0.48\linewidth,clip]%
                   {f+_vs_q2_mud0050_ms0080.eps}
   \hspace{3mm}
   \includegraphics[angle=0,width=0.48\linewidth,clip]%
                   {xi_vs_q2_mud3_ms6.phys.eps}
   \caption{
     Interpolation of $f_+(q^2)$ (left panel) and $\xi(q^2)$ (right panel)
     as a function of $q^2$.
   }
   \label{fig:q2_dep:f+_xi}
\end{center}
\vspace{-5mm}
\end{figure}

We plot our results for $f_0(q^2)$ as a function of $q^2$ 
in Fig.~\ref{fig:q2_dep:f0}.
In this study, 
we simulate small values of $|q^2|$ by using TBCs 
to precisely determine $f_+(0)(\!=\!f_0(0))$. 
The data show small curvature in our region of $q^2$
and are well described by any of the following parametrization forms 

\bea
   f_0(q^2) 
   = 
   f_0(0)(1+c_0 q^2),
   \hspace{5mm}
   f_0(q^2) 
   = 
   f_0(0)(1+c_0 q^2+c_1 q^4), 
   \hspace{5mm}
   f_0(q^2) 
   = 
   \frac{f_0(0)}{1-q^2/M_{\rm pole}^2},
   \hspace{2mm}
   \label{eqn:q2-dep:f0}
\eea
which have also been used in the analyses of experimental data.
In this preliminary report,
we determine the normalization $f_0(0)$ and its slope $f^\prime_0(0)$ 
using the pole form. 
The uncertainty due to this parametrization
is estimated by the largest deviation 
among the results of the above three interpolations.
We note that this uncertainty is similar to or
smaller than the statistical error.

The situation is similar for $f_+(q^2)$, 
which is plotted in the left panel of Fig.~\ref{fig:q2_dep:f+_xi}. 
We observe that $f_+(0)$ is in good agreement with $f_0(0)$ as expected, 
while the latter has slightly smaller uncertainty
due to a better control of its interpolation to $q^2\!=\!0$ 
with the accurate data of $f_0(q^2_{\rm max})$.
We therefore use $f_0(0)$ as the normalization of the vector form factor
in the following analysis.

We plot $\xi(q^2)$ as a function of $q^2$ 
in the right panel of Fig.~\ref{fig:q2_dep:f+_xi}. 
Our data show a very mild dependence on $q^2$
with our statistical accuracy of $\lesssim 30$\,\%.
Note also that 
the leading order (LO) analytic terms in the chiral expansion of $\xi(q^2)$
are independent of $q^2$, 
since $\xi(q^2)$ vanishes as $\propto \! m_s-m_{ud}$ 
in the $SU(3)$ symmetric limit~\cite{kl3:ChPT:f2}.
We interpolate $\xi(q^2)$ to $q^2\!=\!0$ using a linear fit.


\begin{figure}[t]
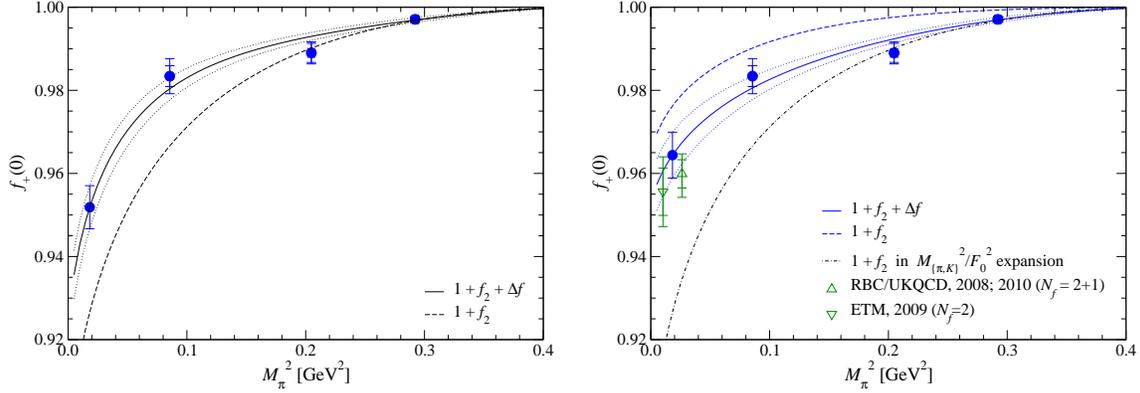

\begin{center}
   \includegraphics[angle=0,width=0.48\linewidth,clip]%
                   {f+_q0_vs_Mpi2.su3.w_F0.nlo+nnlo-anly.eps}
   \hspace{3mm}
   \includegraphics[angle=0,width=0.48\linewidth,clip]%
                   {f+_q0_vs_Mpi2_su3_nlo+nnlo-anly_w-Fpi.eps}
   \caption{
     Chiral extrapolations of $f_+(0)$. 
     The left and right panels show extrapolations 
     based on the chiral expansion
     in terms of $M_{\{\pi,K\}}^2/F_0^2$ and $M_{\{\pi,K\}}^2/F_\pi^2$, respectively.
     In the right panel,
     we also plot $f_+(0)$ from recent calculations
     in $N_f\!=\!2+1$~\cite{kl3:Nf3:RBC/UKQCD} 
     and $N_f\!=\!2$~\cite{kl3:Nf2:ETM} QCD.
   }
   \label{fig:chiral_fit:f+0}
\end{center}
\vspace{-5mm}
\end{figure}

\section{Chiral extrapolation}

Figure~\ref{fig:chiral_fit:f+0} shows our chiral extrapolations of $f_+(0)$ 
based on $SU(3)$ ChPT. 
In this preliminary report,
we employ the fitting form
\bea
   \hspace{-3mm}
   &&
   f_+(0) 
   = 1 + f_2 + \Delta f,
   \hspace{5mm}
   f_2 
   = 
   \frac{3}{2} \left( H_{K\pi} + H_{K\eta} \right),
   \label{eqn:chiral_fit:f+0:1}
   \\
   \hspace{-3mm}
   &&
   H_{PQ} 
   = 
   -\frac{M_P^2+M_Q^2}{128\pi^2F_0^2}
   \left( 1 + \frac{2 M_P^2 M_Q^2}{M_P^4-M_Q^4}
          \ln\left[ \frac{M_Q^2}{M_P^2} \right] \right),
   \hspace{4mm}
   \Delta f
   = 
   \left( \frac{M_K^2-M_\pi^2}{F_0^2} \right)^2
   \left\{c_0 + c_1\frac{M_K^2+M_\pi^2}{F_0^2} \right\},
   \label{eqn:chiral_fit:f+0:2}
   \hspace{9mm}
\eea
where $f_2$ represents the next-to-leading order (NLO) 
contribution~\cite{kl3:ChPT:f2},
and the LO relation $M_\eta^2\!=\!(4M_K^2-M_\pi^2)/3$ 
is used to evaluate $H_{K\eta}$.
We also include the higher order analytic correction $\Delta f$
with $c_{\{0,1\}}$ treated as fit parameters.
The Ademollo-Gatto theorem~\cite{kl3:AG_theorem}
guarantees that 
$f_2$ consists only of the chiral logarithms
with the single low-energy constant (LEC) $F_0$,
that is the decay constant in the chiral limit of three flavors.
Note that Eqs.~(\ref{eqn:chiral_fit:f+0:1}) and (\ref{eqn:chiral_fit:f+0:2})
can be considered 
as an expansion in terms of $M_{\{\pi,K\}}^2/F_0^2$.

We obtain the extrapolation in the left panel of Fig.~\ref{fig:chiral_fit:f+0}
by using $F_0\!=\!52.5(5.1)_{\rm stat}$ 
determined from our study of 
the meson decay constants~\cite{Spectrum:Nf3:JLQCD}.
The convergence of the chiral expansion at the physical quark mass is 
$f_+(0) \! = 1 - 0.073 [f_2] + 0.025(6)\, [\Delta f]$
in contrast to the conventional wisdom that 
$f_2$ is only few percent correction and $\Delta f$ is even smaller.
This is because our estimate of $F_0$ is significantly smaller than
the phenomenological value $F_0\!=\!87.7$~MeV\cite{F0:ChPT:ABT}
and enhances the chiral corrections $\propto \! M_{\{\pi,K\}}^{2n}/F_0^{2n}$.
Note, however, that 
the phenomenological estimate involves large $N_c$ assumptions, 
which are not consistent 
with recent experimental data of the $K_{l4}$ decays~\cite{F0:ChPT:BJ}.

The convergence can be improved by switching the expansion
parameter to $M_{\{\pi,K\}}^2/F_\pi^2$ 
as we already demonstrated in our study of 
$M_\pi$ and $F_\pi$ in $N_f\!=\!2$ QCD~\cite{Spectrum:Nf2:JLQCD}.\footnote{
   See also Ref.~\cite{ReChPT} for a resummation of effects of 
   sea strange quarks, which could be a source of the large deviation
   between $F_\pi$ and $F_0$.
}
The right panel of Fig.~\ref{fig:chiral_fit:f+0} shows the extrapolation
using Eqs.~(\ref{eqn:chiral_fit:f+0:1})\,--\,(\ref{eqn:chiral_fit:f+0:2})
rewritten in terms of $M_{\{\pi,K\}}^2/F_\pi^2$.
We obtain a more convergent expansion
$f_+(0) \ = \ 0.964(6) \ = \ 1 - 0.023 - 0.013(6)$
without a subtle cancellation between $f_2$ and $\Delta f$.
Note also that recent calculations 
in $N_f\!=\!2+1$~\cite{kl3:Nf3:RBC/UKQCD} 
and $N_f\!=\!2$~\cite{kl3:Nf2:ETM} QCD
are consistent with this extrapolation.

\begin{figure}[t]
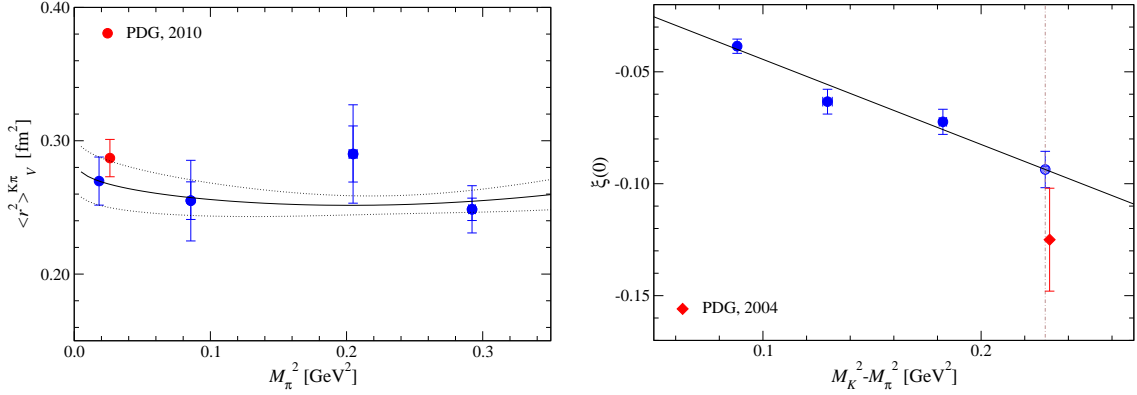

\begin{center}
   \includegraphics[angle=0,width=0.48\linewidth,clip]%
                   {r2_v_vs_Mpi2.eps}
   \hspace{3mm}
   \includegraphics[angle=0,width=0.48\linewidth,clip]%
                   {xi_vs_MK2-Mpi2.eps}
   \vspace{-2mm}
   \caption{
      Chiral extrapolation of $\crad^{K\pi}_V$ (left panel) and 
      $\xi(0)$ (right panel).
   }
   \label{fig:dble_rat:r2_xi}
\end{center}
\vspace{-5mm}
\end{figure}

Our chiral extrapolation of the normalized slope 
$\crad^{K\pi}_V\!=\!6f_+^{\prime}(0)/f_+(0)$
is shown in the left panel of Fig.~\ref{fig:dble_rat:r2_xi}.
We employ the NLO expression~\cite{kl3:ChPT:f2}
\bea
   \crad^{K\pi}_V 
   & = & 
   \frac{12 L_9^r }{F_\pi^2}
   - \frac{3}{64\pi^2 F_\pi^2}
   \left\{  h\left( \frac{M_\pi^2}{M_K^2} \right)
          + h\left( \frac{M_\eta^2}{M_K^2} \right)
          + \frac{2}{3} \ln\left[ \frac{M_\pi^2}{\mu^2} \right] 
          + \frac{5}{3} \ln\left[ \frac{M_K^2}{\mu^2} \right] 
          + \ln\left[ \frac{M_\eta^2}{\mu^2} \right] 
   \right\},
   \hspace{9mm}
   \\
   h(x) 
   & = &
     \frac{x^3-3x^2-3x+1}{2(x-1)^3} \ln\left[ x \right]
   + \frac{1}{2} \left( \frac{x+1}{x-1} \right)^2
   - \frac{1}{3},
   \hspace{5mm}
   \mu = M_\rho,
\eea
plus a higher order analytic correction.
Note that 
$\crad^{K\pi}_V$ has the NLO analytic term with a LEC $L_9^r$
in contrast to $f_+(0)$.
As shown in Fig.~\ref{fig:dble_rat:r2_xi},
our data are well described by this form 
and the extrapolated value is 
in good agreement with the experiment~\cite{PDG:2010}.
Our estimate $L_9^r \times 10^3\!=\!4.1(3)$ 
is slightly smaller than a phenomenological estimate 5.9(4)~\cite{L9:BT},
though the error of our preliminary result is statistical only.

The right panel of Fig.~\ref{fig:dble_rat:r2_xi} shows 
our results for $\xi(0)$ 
as a function of the $SU(3)$ breaking parameter $M_K^2-M_\pi^2$.
In this preliminary analysis,
we parametrize the quark mass dependence of $\xi(0)$ by a simple linear form
\bea
   \xi(0)
   & = & 
   d_0 + d_1 (M_K^2-M_\pi^2),
   \label{eqn:chiral_fit:xi}   
\eea
which is motivated from the ChPT expression of the leading analytic terms 
$\propto M_K^2-M_\pi^2$~\cite{kl3:ChPT:f2}.
Our data are well fitted to this form 
as shown in Fig.~\ref{fig:dble_rat:r2_xi}. 
We obtain $d_0\!=\!-0.006(8)$ confirming that $\xi(0)$ vanishes 
in the $SU(3)$ symmetric limit, as expected.
The extrapolation to the physical point yields 
$\xi(0)\!=\!-0.094(8)$ 
which is consistent with the experimental value 
$-0.125(23)$~\cite{PDG:2004}.


\section{Summary}
\label{sec:summary}


In this article,
we report on our calculation of the kaon semileptonic form factors.
Their normalizations and slopes at $q^2\!=\!0$ are precisely calculated
by using the all-to-all propagator and TBCs. 
We observe a good consistency of $\crad_V^{K\pi}$ and $\xi(0)$ 
with experimental results.

The choice of the expansion parameter is an important issue 
on the convergence of the chiral expansion of $f_+(0)$.
We note that the small value of $F_0$ also enhances
the chiral correction to other observables,
such as the pion and kaon charge radii~\cite{prev1}.
The large deviation $F_\pi\!-\!F_0$ can mainly come from 
effects of sea strange quarks, which are difficult to estimate
solely from experimental data.
A more precise determination of $F_0$ is an important task of lattice QCD
to assess the reliability of the chiral expansion based on $SU(3)$ ChPT.


Our calculations at different values of $m_{ud}$ and $m_s$ are underway
to study systematics of the chiral extrapolations.
It is also interesting to apply two-loop ChPT formulae to our data.
Although the formulae are complicated, 
exact chiral symmetry forbids additional terms due to finite lattice spacings
and provides us a theoretically clean comparison at the higher order.


\vspace{3mm}

Numerical simulations are performed on Hitachi SR11000 and 
IBM System Blue Gene Solution 
at High Energy Accelerator Research Organization (KEK) 
under a support of its Large Scale Simulation Program (No.~09/10-09).
This work is supported in part by 
the Grants-in-Aid for Scientific Research 
(No.~21674002, 21684013, 23105710), 
the Grant-in-Aid for Scientific Research on Innovative Areas
(No. 2004: 20105001, 20105002, 20105003, 20105005), 
and the HPCI Strategic Program 
of the Ministry of Education, Culture, Sports, Science and Technology.


\vspace{-1mm}


\begin{thebibliography}{99}

\bibitem{prev0}
S.Aoki {\it et al.} (JLQCD and TWQCD Collaborations),
Phys. Rev. D {\bf 80}, 034508 (2009).
\vspace{-1mm}

\bibitem{prev1}
T.~Kaneko {\it et al.} (JLQCD Collaboration),
PoS {\bf Lattice 2010}, 146 (2010).
\vspace{-1mm}

\bibitem{A2A} 
G.S.~Bali {\it et al.} (SESAM Collaboration),
Phys. Rev. D {\bf 71}, 114513 (2005);
J.~Foley {\it et al.} (TrinLat Collaboration),
Comput. Phys. Commun {\bf 172}, 145 (2005).
\vspace{-1mm}

\bibitem{TBC}
P.F.~Bedaque, 
Phys. Lett. B {\bf 593}, 82 (2004).
\vspace{-1mm}

\bibitem{kl3:drat:1}
D.~Be\'cirevi\'c {\it et al.}, 
Nucl. Phys. B {\bf 705}, 339 (2005).
\vspace{-1mm}

\bibitem{kl3:drat:2}
N.~Tsutsui {\it et al.} (JLQCD Collaboration), 
PoS {\bf LAT2005}, 357 (2005);
C.~Dawson {\it et al.} (RBC Collaboration)
Phys. Rev. D {\bf 74}, 114502 (2006).
\vspace{-1mm}

\bibitem{kl3:ChPT:f2}
J.~Gasser and H.~Leutwyler,
Nucl. Phys. B {\bf 250}, 517 (1985).
\vspace{-1mm}

\bibitem{kl3:Nf3:RBC/UKQCD}
P.A.~Boyle {\it et al.} (RBC and UKQCD Collaborations),
Phys. Rev. Lett. {\bf 100}, 141601 (2008).
\vspace{-1mm}
 
\bibitem{kl3:Nf2:ETM}
V.~Lubicz {\it et al.} (ETM Collaboration), 
Phys. Rev. D {\bf 80}, 111502 (2009).
\vspace{-1mm}

\bibitem{kl3:AG_theorem}
M.~Ademollo and R.~Gatto,
Phys. Rev. Lett. {\bf 13}, 264 (1964).
\vspace{-1mm}

\bibitem{Spectrum:Nf3:JLQCD}
J.~Noaki {\it et al.} (JLQCD and TWQCD Collaborations),
PoS {\bf Lattice 2010}, 117 (2010).
\vspace{-1mm}

\bibitem{F0:ChPT:ABT}
G.~Amoros, J.~Bijnens and P.~Talavera,
Nucl. Phys. B {\bf 602}, 87 (2001).
\vspace{-1mm}

\bibitem{F0:ChPT:BJ}
J.~Bijnens and I.~Jemos,
arXiv:1103.5945 [hep-ph].
\vspace{-1mm}

\bibitem{Spectrum:Nf2:JLQCD}
J.~Noaki {\it et al.} (JLQCD and TWQCD Collaborations),
Phys. Rev. Lett. {\bf 101}, 202004 (2008).
\vspace{-1mm}

\bibitem{ReChPT}
S.~Descotes-Genon {\it et al.},
Eur. Phys. J. C {\bf 52}, 141 (2007).
\vspace{-1mm}

\bibitem{PDG:2010}
K.~Nakamura {\it et al.} (Particle Data Group),
J. Phys. G {\bf 37}, 075021 (2010).
\vspace{-1mm}

\bibitem{L9:BT}
J.~Bijnens and P.~Talavera, 
JHEP {\bf 03}, 046, (2002).
\vspace{-1mm}

\bibitem{PDG:2004}
S.~Eidelman {\it et al.} (Particle Data Group),
Phys. Lett. B {\bf 592}, 1 (2004).

\end{thebibliography}
\end{document}